# Thermal conductivity of macroporous graphene aerogel measured using high resolution comparative infrared thermal microscopy

Jasmine M. Cox[a*], Jessica J. Frick[a], Chen Liu[c], Zhou Li[b,d], Yaprak Ozbakir[d], Carlo Carraro[d], Roya Maboudian[d], Debbie G. Senesky[a,b]

*Corresponding Author: jascox@stanford.edu

[a]Department of Electrical Engineering, Stanford University, California 94305, United States, [b]Department of Aeronautics & Astronautics, Stanford University, California 94305, United States, [c]Department of Mechanical Engineering, Stanford University, California 94305, United States, [d]Department of Chemical and Biomolecular Engineering, and Berkeley Sensor & Actuator Center, University of California, Berkeley, California, 94720, United States

**Abstract**

Graphene aerogel (GA) is a promising material for thermal management applications across many fields due to its lightweight and thermally insulative properties. However, standard values for important thermal properties, such as thermal conductivity, remain elusive due to the lack of reliable characterization techniques for highly porous materials. Comparative infrared thermal microscopy (CITM) is an attractive technique to obtain thermal conductance values of porous materials like GA, due to its non-invasive character, which requires no probing of, or contact with, the often delicate structures and frameworks. In this study, we improve upon CITM by utilizing a higher resolution imaging setup and reducing the need for pore-filling coating of the sample (previously used to adjust for emissivity). This upgraded setup, verified by characterizing porous silica aerogel, allows for a more accurate confirmation of the fundamental thermal conductivity value of GA while still accounting for the thermal resistance at material boundaries. Using this improved method, we measure a thermal conductivity below 0.036 W/m·K for commercial GA using multiple reference materials. These measurements demonstrate the impact of higher resolution thermal imaging to improve accuracy in low density, highly porous materials characterization. This study also reports thermal conductivity for much lower density (less than 15 mg/cm$^3$) GA than previously published studies while maintaining the robustness of the CITM technique.

**Statements and Declarations:** This work was supported by the National Science Foundation [grant numbers 1929363, 1929447]; Part of this work was performed at the Stanford Nano Shared Facilities (SNSF), supported by the National Science Foundation [award number ECCS-2026822].

**Acknowledgements:** The authors would like to thank Prof. Kenneth Goodson and Prof. Mehdi Asheghi and the members of Stanford NanoHeat Lab at Stanford University for usage of their equipment and helpful insights about technique verification. The authors would also like to thank Rachel Ormsby and the team members at external collaborator Redwire Inc. for their helpful suggestions in furthering this work.

## 1. Introduction

Graphene aerogel (GA) represents an emerging class of highly porous materials with potential applications in aerospace, electronics, and structural construction [1,2]. Since the initial discovery of graphene in 2004, many efforts have been made to increase GA synthesis volume and yields[3–5]; however, thermal, mechanical, electrical, and radiative properties remain varied in research. For example, GA's electrical conductivity value of up to ∼1 × $10^2$ S/m is reported based only upon extrapolations of data obtained from GA-like materials typically formed from graphene oxide in a mixture with a framework material to control structure formation during hydrothermal reduction and thus modifying the inherent GA crosslinked structure [6–8]. Other studies, when commenting on the electrical properties of GA, are actually referencing studies on two-dimensional (2D) graphene–completely disregarding the impact its three-dimensional (3D) microstructure has on the materials' electrical conductivity [9–13]. This lack of consensus is due in part to the challenges presented by the stochastic pore distribution structure that is inherent in the production process–namely, hydrothermal reduction of graphene oxide flakes, which induces random crosslinking of 2D graphene sheets to form the 3D framework, and additionally, to the understood limitations of existing characterization models relying on solid or very low porosity materials to uphold theoretical analysis [14–16]. These unique materials remain without a standardized technique for characterization or modeling of thermal properties. Review of the existing literature determines that optical non-invasive probing is underdeveloped but is the most promising type of characterization to determine GA's fundamental properties without compromising it with embedded materials or alterations of the macroporous structure that make most characterization techniques more accurate but give little insight into the intrinsic properties of GA derived from only rGO [17,18]. Reliable characterization of GA is crucial to identifying plausibility in applications, specifically characterizing its thermal conductivity for potential use as thermal insulation of spacecraft vehicles or usage as a semiconducting material in electronics for extreme applications [2,19,20].

In the current literature, there are only two reports on the fundamental thermal transport properties of unmodified GA. One, by Xie et al. in 2016, use the steady-state electro-thermal technique (SET) to measure very low-density GA (~3.9 - 4.2 mg/cm$^3$) with a reported $k_{GA}$ of (4.7 – 5.9) x $10^{-3}$ W/m·K at room temperature (RT) [21]. This technique is limited by the requirement of a standard characterized density and heat capacitance for the SET to be accurately applied [11]. In the second report, Fan et al. use the comparative infrared thermal microscopy (CITM) technique to measure apparent thermal conductivity ($k_{GA}$ ~ 0.12 – 0.36 W/m·K) of higher density GA ($\rho$ ~ 14.1 – 52.4 mg/cm$^3$) to explore the effect of heat treatments on GA thermal properties [22]. While this study establishes a practical implementation of the CITM technique for porous materials, it is limited by the low resolution of the resulting IR thermal images, as well as the lack of viable exploration of reference materials utilized in setup (as the authors note).

This work improves upon the CITM technique as described in Fan et al. by addressing the two limitations noted by the authors. By utilizing an IR microscope designed to account for varied reflectance in a single image, the resolution of images of similar size stacks is greatly increased in comparison to those from an IR camera setup. This increases the accuracy of the measurement by decreasing the representation of thermal resistance at stack boundaries in the image pixels. This further results in more image pixels of data to associate with the actual thermal gradient being achieved in the CITM setup. The technique setup is additionally improved by evaluating a wider range of reference materials with thermal conductivities comparable to GA. In this work, materials with lower $k$ values than the amorphous quartz previously reported on are tested in agreement with the very low thermal conductivity reported by Fan et al. The reported value of k given in that study is especially indicative of GA being more comparable to highly insulating PDMS material that is considered a standard reference material for this technique when applied to materials with expected thermal conductivities in that range to ensure comparable thermal gradients in each stack section [23]. Utilizing reference materials in this higher resolution imaging setup will allow for a clear determination of the thermal conductivity of GA in comparison to a range of reference materials more insulating than amorphous quartz down to PDMS. This work further reports on the thermal conductivity of GA with a moderate density representative of the commercially available GA, both of higher and lower density than that of the two previously reported studies broadening the implication of density and porosity on the overall insulating effect of the material.

## 2. Materials and Methods

2.1 Materials

Graphene aerogel is purchased commercially from Graphene Supermarket. NBK-7 and B270 glass is purchased pre-fabbed in 10 x 10 mm square cross sections of 1-2 mm thickness from Edmund Optics, Inc. The PDMS is purcha sed commercially from Interstate Specialty Products in 1 mm thick sheets that 10 x 10 mm pieces are then cut from.

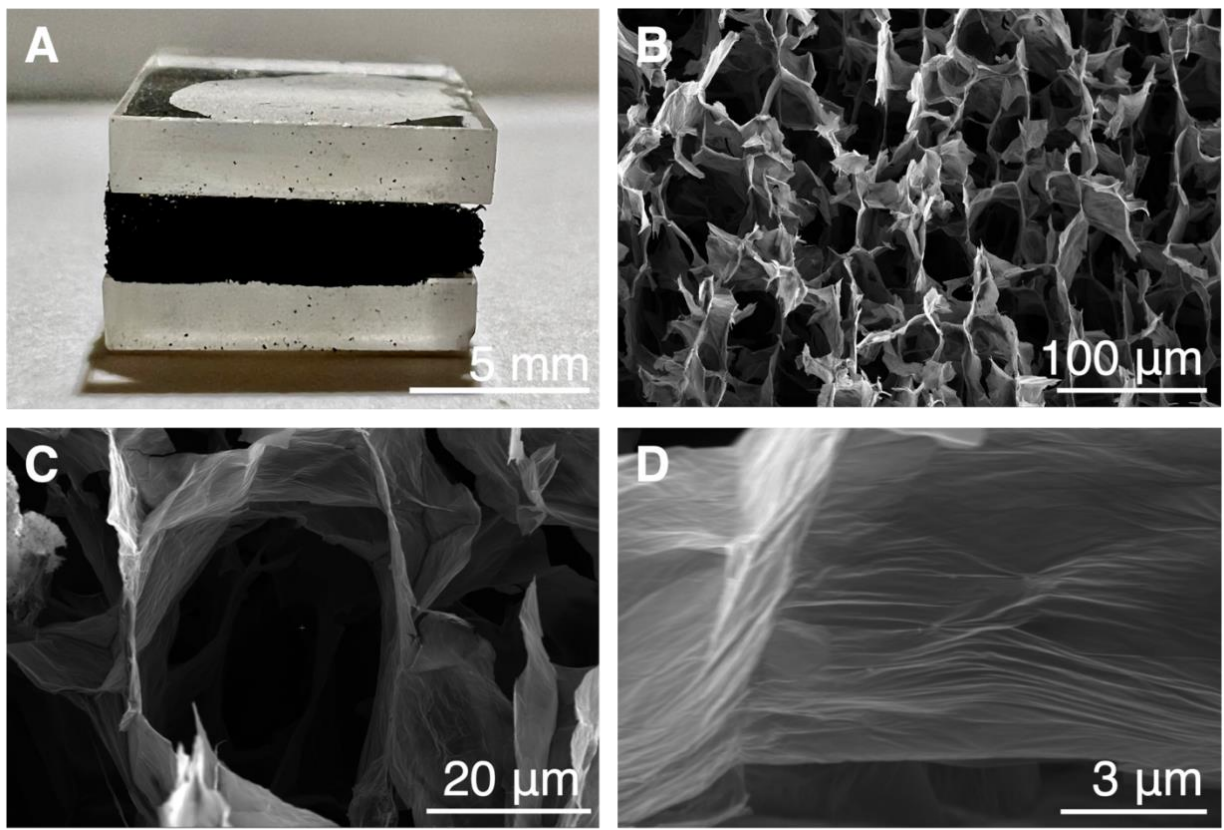

**Fig. 1** *(A) Optical image of stack with graphene aerogel sandwiched between two pieces of 2-mm-thick glass (B) SEM image of GA showing high number of pores (C) SEM image of singular GA pore demonstrating the large pore size (D) SEM image of GA showing the irregular striation pattern of 2D graphene sheets forming GA*

2.2 Morphology

The commercial GA microstructure was imaged by scanning electron microscopy (SEM) using a Thermo Fisher Scientific Apreo S LoVac instrument (Figure 1B-D). SEM was performed at a voltage of 5 kV and a current of 50 pA in secondary electron (SE) imaging mode with a working distance of around 10 mm.

2.3 Thermal Characterization

The GA thermal conductivity is measured using the CITM technique, which was first demonstrated by Fan et al [22]. To prepare a GA sample for measurement, GA was stacked on top of a piece of reference material, either a type of soda-lime glass or PDMS, with another piece of the same reference material placed on top, creating a "stack" (Figure 1A). Between each layer where GA and reference material met, silver paint was applied to reduce thermal resistance and adhere the stack components together. The stack was then placed in a copper holder, making contact on the non-GA side of the reference material, with a heater attached to one side of the holder. The holder contact area with the sample was 10 x 10 mm$^2$, the same size as the in-plane cross section of the stack with respect to heat flux. The unheated side of the holder acted as a copper heat sink. A schematic of the set-up is demonstrated in Figure 2.

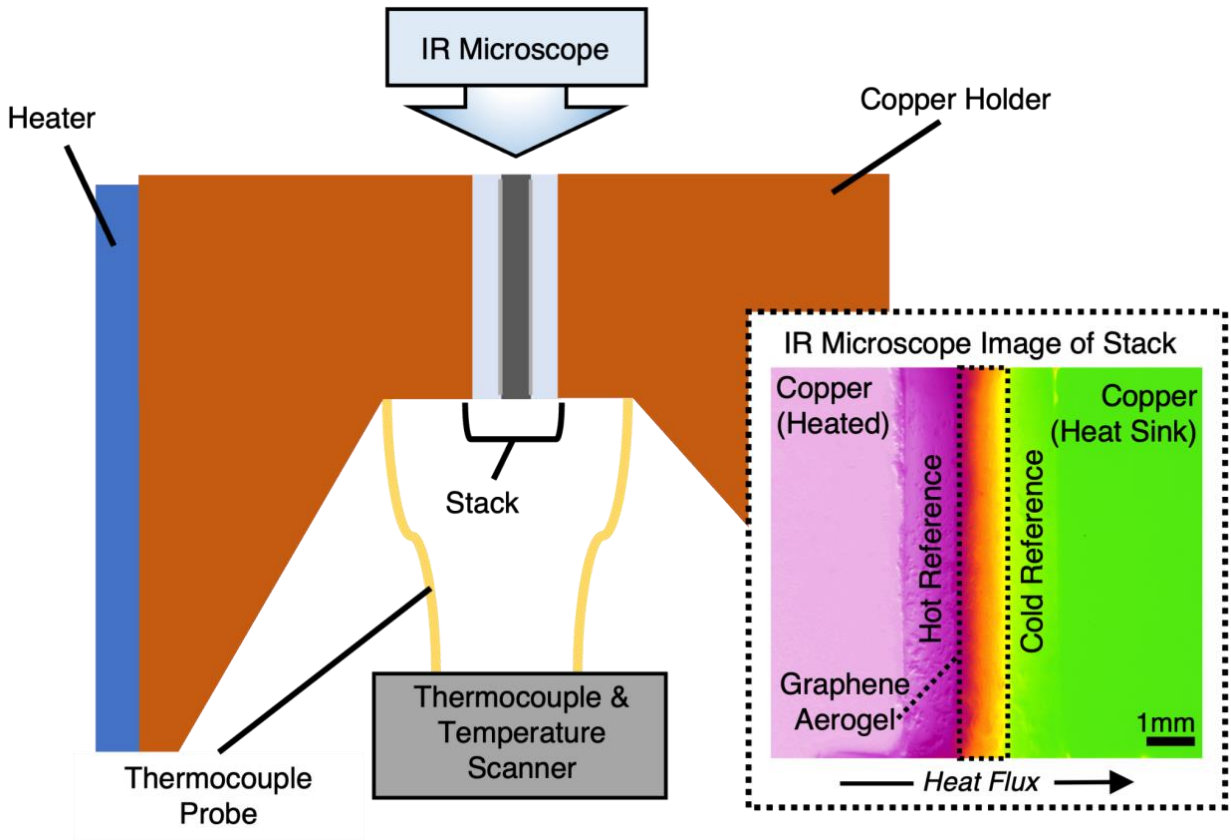

**Fig. 2** *Schematic of structure and set-up used to measure thermal conductivity using the CITM technique. The inset is an annotated example output IR thermal image of a stack containing PDMS as a reference material using this setup*

## 3. Experimental

Two types of soda-lime glass, NBK-7 and B270, were used for this experiment due to their range of low thermal conductivity values, $k$ = 1.14 and 0.92 W/m·K, respectively. PDMS was also used due to its even lower $k$ value, to provide a range of potentially comparable thermally conducting materials to the GA. All components of the stack had a cross section of roughly 10 x 10 mm with the thickness of the GA made comparable to the thickness of each respective reference material, ranging from 0.7 to 2 mm. Maintaining comparable thickness of the GA and reference material is necessary to apply the heat flux equation.

The IR microscope (InfraScopeTM MWIR Temperature Mapping Microscope, Quantum Focus Instruments Corp.) algorithmically corrects for emissivity of varied materials and has a temperature resolution of 0.1 K. Before test imaging, all sample stacks were heated uniformly to 80 °C to provide an emissivity baseline for correction in the exact focus and position used for testing. This practice eliminates the need for applying graphite paint to surfaces facing the MWIR microscope camera, as done in previous CITM studies [24].

During measurements, a heat flux was generated in one direction by heating only one side of the copper holder to 80 °C and allowing the other side to remain ambient while the microscope camera captured the temperature distribution across the plane. The resulting 2D temperature map was then reduced to a 1D distribution by averaging the temperature values perpendicular to the heat flux. Removal of pixel data correlating to temperature values of the copper holder and at the boundaries of each stack component was done to account for thermal resistance. Figure 3 demonstrates a typical temperature-pixel distance profile generated from each measurement with the thermal gradient of each specific stack material calculated from the least-squares fit of the measured temperature data.

Fourier's law applies to the transport of heat through the stack as a temperature differential is achieved based upon understanding of thermal transport in carbon-based materials and applied to this 3D structure [25]. The theoretical verification of this analysis is explored previously by Fan et al [22]. By specifying the same cross-sectional area for each stack component, the steady state heat conduction equation can be applied:

$$q'' = -k_{ref}(dT/dx)_{ref} = -k_{GA}(dT/dx)_{GA} \qquad (1)$$

where $q''$ is the heat flux through the GA sample stack determined from the gradient of temperature $T$ with respect to linear distance $x$. By applying this equation, the thermal conductivity of GA can be extracted from the heat flux derived from the averaged thermal gradients of the reference material and the known thermal conductivity of the reference materials. Using Eq. (1), the thermal conductivity of GA is then derived using the measured thermal gradient of GA.

To verify the techniques robustness, a benchmarking test was completed using the more thermally characterized silica aerogel with PDMS as reference material to produce comparable results to published literature [16,26–28]. Each GA measurement was repeated three times with similar conditions for each reference material and the measured thermal conductivity values were averaged respectively. Results of this verification study are represented in the supporting information, Figure S1.

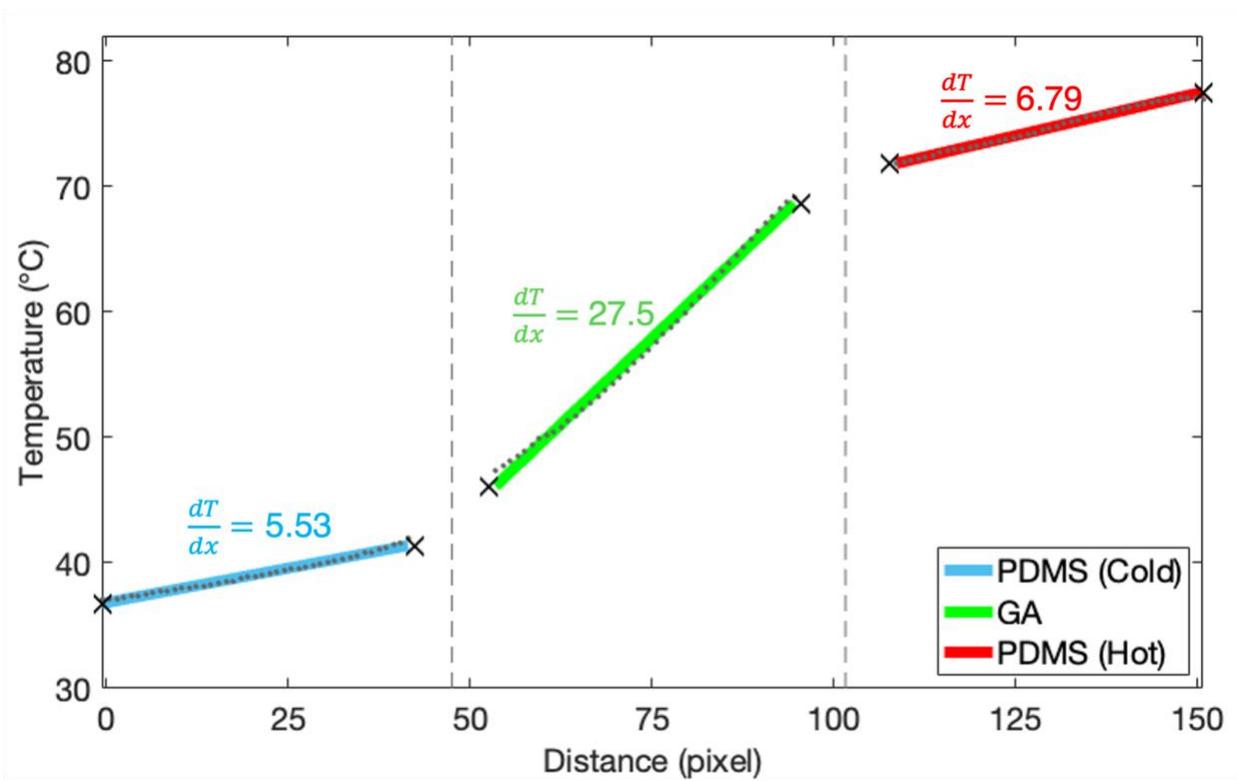

**Fig. 2** *Linear best fit curves of temperature distribution in sample stacks consisting of graphene aerogel between two pieces of PDMS reference material with a scale of 19.9 μm/pixel. The boundary pixels have been removed to account for thermal resistance*

## 4. Results and Discussion

**Table 1.** *Experimental Results Compared to Relevant Literature*

| Publication | Technique | Reference Material | $k_{ref}$ (W/m·K) | GA Density (mg/cm$^3$) | Average $k_{GA}$ (W/m·K) |
|---|---|---|---|---|---|
| *This work* | CITM | NBK7 glass | 1.14 | 12.5 | 0.052 +/- 0.002 |
| | | B270 glass | 0.92 | | 0.051 +/- 0.013 |
| | | PDMS | 0.16 | | 0.033 +/- 0.003 |
| Fan et al. [22] | CITM | Quartz | 1.3 | 33.3 | 0.24 |
| Xie et al. [21] | SET | – | – | 4.1 | 0.0053 |

A summary of the GA thermal conductivity $k_{GA}$ measurements is shown in Table 1 and compared with the results reported in Ref. 21 and 22. More detailed value reporting can be found in the supporting information, Table S1. The reference material NBK7 ($k$ = 1.14 W/m·K) resulted in an average $k_{GA}$ = 0.052 W/m·K. Using the glass reference sample B270 ($k$ = 0.92 W/m·K) resulted in a comparable $k_{GA}$ range, from 0.0423 to 0.662 W/m·K. To examine the influence of the reference material thermal conductivity on the resulting $k_{GA}$ value, we chose PDMS as a reference sample, which has a $k_{ref}$ substantially lower relative $k_{ref}$ = 0.16 W/m·K. These measurements resulted in a $k_{GA}$ range from 0.0306 to 0.0358 W/m·K. The IR images and graphical analysis of these tests can be found in the supporting information, Figures S2-S10.

The examination of multiple reference materials to measure the thermal conductivity of GA demonstrates the previously reported optical measurements using glass substrates with similar thermal conductivities to NBK7 and B270 glass may be an indication of incompatible reference material to apply the heat flux equation. By recreating similar stack samples with even lower glass thermal conductivities, it is demonstrated that CITM measurements are only applicable in the scenario that the reference materials utilized have a similar thermal conductivity as the GA. Measuring GA with PDMS as a reference material results in a much lower thermal conductivity value than that of the materials with comparable GA density reported in the previous studies. Repeated characterization of new GA samples with PDMS demonstrated a more consistent result than other substrates implying the CITM technique is accurate with porous materials so long as a material with a comparable thermal conductivity is in contact. This result agrees with the fundamental principles assumed by the heat flux equation so long as all materials have a similar cross section and cross-sectional area.

CITM measurement of the thermal conductivity of graphene aerogel maintains the porous structure and allows for more accurate investigation of the fundamental properties. Measurement of GA at a standardized density of 12.5 mg/cm$^3$ using varied reference materials demonstrates that the GA of graphene aerogel is very low and indicates that the aerogel may only be accurately measured while in contact with other insulating materials such as PDMS. The resulting thermal conductivity of GA measured with PDMS as reference indicated that the GA is more comparable to silica aerogel in thermal performance than reliably indicated before.

In previous studies, the CITM technique was limited by specific setup factors. Previous studies utilized imaging that increased the pixel scale to over 100 μm/px resulting in less accurate accounting for thermal boundaries and resistance. These studies typically utilized less than 30 pixels in each plane to optically measure the stack thermal gradient in order to avoid increasing the overall stack size and the heat loss due to convection and measurement error. This study demonstrates a more robust variation of the CITM setup for porous materials where the varied emissivity can be accounted for without using a coating that could potentially result in filled GA pores and altered thermal conductivity. The highly reduced scale of 19.9 μm/px enables for a more exact understanding of the thermal gradient in each material section of the stack and reduces the error introduced by thermal resistance at the material boundaries.

## 6. Conclusions

This work demonstrates an improved CITM thermal characterization technique to measure thermal conductivity in GA with a moderate density found commercially but uncharacterized in literature. The use of higher resolution imaging to obtain thermal gradient values of an image stack reduces the attribution of thermal boundary in the overall determination of material thermal conductivity. GA is demonstrated to have a measured $k \sim 0.03$ W/m·K when compared to PDMS in reference. The results confirm that optical measurement techniques can be more accurately applied to delicate mesostructures without need for filler materials to account for large pores in the material surface and thus alter the thermal performance. It is further confirmed that reference materials have a significant effect on quantifying heat flux, and thus careful consideration of the appropriate comparable reference and sample material with an expected thermal conductivity or thermal gradient remains necessary. The results demonstrate that GA is a highly thermally insulating material and gives verification to previous work implicating that GA is a suitable material for thermal insulation and heat shield applications requiring a lightweight material without losing surface area, while maintaining other fascinating physical characteristics that may be altered to address porosity complications in other measurement techniques.

**CRediT Author Statement:** *Jasmine M. Cox:* Investigation, Writing-Original Draft, Visualization, Formal Analysis, Data Curation, Conceptualization, Methodology. *Jessica J. Frick:* Writing-Review & Editing, Supervision, Conceptualization, Visualization. *Chen Liu:* Conceptualization, Methodology. *Zhou Li:* Conceptualization, Supervision. *Yaprak Ozbakir:* Conceptualization *Carlo Carraro:* Project administration, Conceptualization, Writing-Review & Editing. *Roya Maboudian:* Project administration, Funding acquisition, Conceptualization, Writing-Review & Editing**.** *Debbie G. Senesky:* Project administration, Funding acquisition, Supervision, Conceptualization.